\documentclass[fleqn,twoside]{article}
\usepackage{espcrc2}

\usepackage{graphicx}
\usepackage[figuresright]{rotating}

\def\be{\begin{eqnarray}&&}
\def \nonu {\nonumber \\&&}
\def\psla{\slash \! \! \!}
\def\Psla{\slash \! \! \! \! }

 \def\ee{ \end{eqnarray}}

\newcommand{\AmS}{{\protect\the\textfont2
  A\kern-.1667em\lower.5ex\hbox{M}\kern-.125emS}}

\title{A covariant investigation of  Neutral Vector Mesons: \\ dynamical properties 
and electromagnetic  decay widths
}
\author{Tobias Frederico \address{ITA, 12.228-900, S\~ao Jos\'e dos
Campos, S\~ao  Paulo, Brazil}, Emanuele Pace \address
{Universit\`a di Roma  "Tor Vergata" and Istituto Nazionale di Fisica Nucleare
- Sezione di "Tor Vergata", Via della Ricerca Scientifica 1, 00133 Rome, Italy}, 
Silvia Pisano \address{Institut de Physique Nucl\'eaire d'Orsay, 91404
Orsay, France}, 
Giovanni Salm\`e \address{Istituto Nazionale di Fisica Nucleare
- Sezione di Roma \\ 
        P.le A. Moro 2, 00185 Rome, Italy}}
\begin{document}
\begin{abstract}
 A simple, but   fully-covariant model  
 for describing neutral Vector
 Mesons, in both   
light and  heavy sectors, is briefly illustrated. 
The main ingredients of our relativistic constituent model 
are i)
an Ansatz for the Bethe-Salpeter vertex for 
Vector Mesons, 
and ii) a Mandelstam-like formula for the electromagnetic decay widths. 
The free parameters of our approach are fixed through a comparison
with
the valence transverse-momentum 
distribution, $n(k_\perp)$, obtained within  phenomenological 
 Light-Front Hamiltonian Dynamics models reproducing the mass spectra.
 Preliminary results for  the transverse-momentum distributions, 
 the parton distribution and     
the electromagnetic decay constants are shown.
\vspace{1pc}
\end{abstract}

\maketitle
\section{INTRODUCTION}
Aim of this contribution  is to illustrate  a fully
covariant model for describing the  electromagnetic 
(em) decay of neutral 
 Vector Mesons (VM), in both light and heavy sectors (see \cite{phipsi} for a
 preliminary presentation of our approach). For the present stage, 
 we have considered only the ground states, to be described  by using  
   Bethe-Salpeter 
(BS)
amplitudes  with a simple analytic form, namely exhibiting only single poles. This
assumption allows us 
 to easily perform the analytic integration needed for  
  calculating the constituent quark (CQ)  transverse-momentum distribution 
  inside the
  VMs
(see, e.g. Refs. \cite{melo02} and \cite{pion09} for the pion case) 
and then
em decay widths. The free parameters present in  the Ansatz for each VM are fixed
by comparing  
the so-called {\em  valence transverse-momentum distribution}, $n(k_\perp)$, 
with the same quantity
evaluated within  a Light-Front
 Hamiltonian Dynamics (LFHD) approach exploiting the eigenfunctions
 of two phenomenological mass operators, presented in Refs. \cite{qcd_tob} and
 \cite{GI}, and able to reproduce the VM spectra. It should be pointed out
 that the free parameters represent an effective way for  including 
some non perturbative features in our analytical Ansatz. 

A possible form of
the BS amplitude for an interacting $q\bar q$ system with $J=1$, can be 
written
as follows
\be
\Psi_\lambda(k,P)=S(k,m_1)~{\cal V}_{\lambda}(P)~S(k-P,m_2)\times \nonu
\Lambda_{VM}(k,P)
\label{bsa}
\ee
where $S(k,m)=
S(k_{on},m) + {\gamma^+ / (2 k^+)}$ is the Dirac propagator 
of a constituent quark with mass $m$, $k^\mu_{on}\equiv\{k^-_{on},k^+,{\bf k}_\perp\}$
with 
$k^-_{on}=(m^2+|{\bf k}_\perp|^2)/k^+$, 
 $P^\mu$ the  four-momentum of a VM
 with  mass $P^2=M_{VM}^2$, $\Lambda_{VM}(k,P)$ the
momentum dependence of the BS amplitude, and ${\cal V}_{\lambda}(P)=\epsilon_\lambda(P)
\cdot V(P)$ with $\epsilon^\mu_\lambda(P)$ the polarization
four-vector ($\lambda$ is the helicity) and   $V^\mu(k,k-P)$
 the Dirac structure given by  the following 
familiar expression, transverse to $P^\mu$, (see, e.g., \cite{ji92})
\be 
V^\mu(P) ={M_{VM} \over M_{VM}+m_1+m_2}~\left [\gamma^\mu -
{P^\mu ~\Psla P \over M_{VM}^2} + \right. \nonu \left. + ~ \imath ~ {1 \over M_{VM}}~
\sigma^{\mu\nu} P_\nu\right ]
\ee
Then, the  scalar product  ${\cal V}_{\lambda}(P)$ reduces to
\be
{\cal V}_{\lambda}(P)={\psla  \epsilon_\lambda\left [M_{VM} -
 {\Psla P }\right ]\over M_{VM}+m_1+m_2}
 \label{dstruc}
\ee
It should be pointed out that one could have two different $\Lambda_{VM}(k,P)$'s, one 
 for each Dirac
structure  in ${\cal V}_{\lambda}(P)$ (i.e. 
$\psla  \epsilon_\lambda$ and $\psla  \epsilon_\lambda {\Psla P }$), 
but at the present stage we assume  a
simpler form, with the same $\Lambda_{VM}(k,P)$ for the two structures, that 
  leads to the expected Melosh Rotation
factor for a
$^3S_1$ system, in the limit of non interacting systems, i.e. $M_{VM}\to M_{free}$
or $P^\mu \to P^\mu_{free}$
 \cite{jaus91}. Therefore the following comparison with the
 phenomenological LFHD outcomes can be  made more strict. 
 
 In the 
preliminary calculations presented in this contribution,
the momentum dependence of the BS amplitude is given by  the following    
Ansatz, with only single poles,  viz
\be \Lambda_{VM}(k,P)= {\cal  N}~\nonu
\left[k^2-m_{1}^2+(P-k)^2-m_{2}^2\right ]~\Pi_{i=1,3} \times \nonu  
{1\over \left[k^2-m_{R_i}^2 +\imath\epsilon\right ]
\left [(P-k)^2-m_{R_i}^2 +\imath\epsilon\right ]}
\label{lambda}
\ee
where $m_{R_i}$, $i=1,2,3$ are  free parameters (determined as
described below), and ${\cal  N}$ the normalization factor, that is obtained
 by imposing the  normalization
  for the BS amplitude in 
Impulse Approximation, i.e. by adopting    constituent quark free propagators.
 The form chosen for $\Lambda_{VM}(k,P)$ allows one: i) to implement   
 the correct symmetry under the exchange of the quark momenta
(for equal mass constituents), as follows from the charge conjugation of the
neutral mesons; ii) to regularize the integrals needed in our approach for the evaluation of valence wave
functions, decay constants, transverse-momentum distributions, etc.; and iii) 
to avoid
 any free propagation in the valence wave function, given the  presence of the
 numerator in 
 $\Lambda_{VM}(k,P)$ (see also \cite{phipsi,Gross}). 

To determine $m_{R_i}$ in Eq. (\ref{lambda}),  we have introduced  the 
  transverse-momentum
distribution of a CQ inside the VM, adopting the same definition
already applied  by de Melo 
et al. \cite{melo02} for a covariant description of 
 the pion. In a frame where ${\bf P}_\perp={\bf 0}$, one  first 
defines the
valence component of the VM state in the standard way 
(see, e.g., \cite{HK}), viz.
\be
\Psi^{val}_{VM}(\xi,{\bf k}_\perp; m_{R_i})=\int dk^-~S(k_{on},m_1)~
\times\nonu
{\cal V}_{\lambda}(P)~
\Lambda_{VM}(k,P)S\left((k-P)_{on},m_2\right )
\label{val}\ee
with $\xi=k^+/P^+$.  
The normalization factor in $\Lambda_{VM}(k,P)$ (cf  Eq. (\ref{lambda})) is
evaluated  
 in Impulse
Approximation by exploiting the total-momentum sum rule, that reads for the plus
component as follows
\be 
 P^+ =~{N_c\over 2}
 \int\frac{d^4k}{(2\pi)^4}~{\bar\Lambda_{VM}(k,P)\over 
  \left[(k-P )^2- m^2_2+\imath\epsilon\right]}~
~\times \nonu 
 { { \bf T}^+_{\lambda,\lambda}(P,k) \over \left[(k-P)^2- m^2_2+\imath\epsilon\right]}
 { \Lambda_{VM}(k,P)\over (k^2 - m^2_1+\imath \epsilon) }
 \ee
with $N_c=3$ and 
\be
{ \bf T}^+_{\lambda,\lambda}(P,k)=Tr\left \{{\cal V}_{\lambda}(P)
~\times \right .\nonu \left.({\psla k}-\psla{P}  +m_2)
\gamma^{+} ~(\psla{k}-\psla{P}+m_2) ~\times \right .\nonu \left.
 {\cal V}^\dagger_{\lambda}(P)~(\psla{k}+m_1)\right \}=
\nonu= -2 (P^+-k^+)\left[\phantom{{(k^- -k^-_{on}) \over 2}}\hspace{-1.5cm}
{\cal T}^+(P,k,on)+\right. \nonu \left.
{(k^- -k^-_{on}) \over 2}{ \cal T}^+(P,k,ist)\right]
\label{trace}\ee
In Eq. (\ref{trace}), 
$
{\cal T}^+(P,k,on)$ 
and 
${ \cal T}^+(P,k,ist)$ are the on-shell and the instantaneous
contributions, respectively, according to the well-known decomposition of the Dirac
propagator (see below Eq. (\ref{bsa})).
Their explicit expressions will be given elsewhere \cite{fpps10}.
 Notice that, after performing the  trace, the dependence upon the
helicity disappears, and therefore its presence in ${ \bf
T}^+_{\lambda,\lambda}(P,k)$ becomes dummy. 
\begin{table*}[tbh]
\caption{The mass eigenvalues for the ground states of the neutral VMs investigated in this contribution 
compared with the
corresponding experimental values. The   mass
operator corresponds to the  one in Ref. \cite{qcd_tob} (see \cite{fpps10} for
details). The  quark masses adopted for the actual evaluation are also listed.}
\label{tab1}
\vspace{.3cm}
\hspace{3.8cm}\begin{tabular}{cccc}\hline
{VM} & $m_q (MeV)$ &$M^{th}_{VM}$ (MeV) & $M^{exp}_{VM}$ (MeV)
\\
\hline
$\rho$ &328 &777.0&775.500 $\pm$ 0.4   
\\  
$\phi$ &449& 1020.7 &1019.455 $\pm$ 0.020
\\  
$J/\psi$& 1559&  3082.9 & 3096.916 $\pm$ 0.011   
\\  
$\Upsilon$ &4891& 9537.2 &9460.300 $\pm$ 0.268   
\\  \hline \\
\end{tabular} 
\end{table*}
\begin{table*}[tbh]
\caption{The same as in Table 1, but for  the   mass
operator of Ref. \cite{GI}.}
\label{tab2}
\vspace{.3cm}
\hspace{.5cm}\begin{tabular}{cccc}\hline
{VM} & $m_q (MeV)$ &$M^{th}_{VM}$ (MeV) & $M^{exp}_{VM}$ (MeV)
\\
\hline
$\rho$ &220 &777&775.500 $\pm$ 0.4   
\\  
$\phi$ &419& 1016 &1019.455 $\pm$ 0.020
\\  
$J/\psi$& 1628&  3091 & 3096.916 $\pm$ 0.011   
\\  
$\Upsilon$ & 4977& 9460 &9460.300 $\pm$ 0.268   
\\  \hline
\end{tabular} 
\end{table*}
Finally, let us introduce  the  valence transverse-momentum distribution, $n_{val}(k_\perp)$,
 given by
\be
n_{val}(k_\perp)={N_c\over (2\pi)^3\left [P^+\right]^2 P^{VM}_{q\bar q}}
\int_0^{2\pi}d\theta_{k_\perp} ~\times \nonu \int_0^1 d\xi{ {\cal T}^+(P,k,on)
\over
\xi ~(1-\xi)}~|\Phi^{val}_{VM}(\xi,{\bf k_{\perp}};
m_{R_i})|^2
\label{nk}\ee 
where $P^{VM}_{q\bar q}$ is the probability of the valence component, 
given by
\be
P^{VM}_{q\bar q}=
{N_c\over (2\pi)^3\left [P^+\right]^2}
\int_0^1 {d\xi \over
\xi ~(1-\xi)}~\times \nonu\int d{\bf k}_\perp ~{\cal T}^+(P,k,on)~
|\Phi^{val}_{VM}(\xi,{\bf k_{\perp}};
m_{R_i})|^2
\label{nvm}\ee
and 
\be
\Phi^{val}_{VM}(\xi,{\bf k_{\perp}};
m_{R_i})=~k^+~(P-k)^+~\int dk^-~\times \nonu
{\Lambda_{VM}(k,P) \over (k^2-m^2_1)~\left[(k-P)^2-m^2_2\right]}
\ee
  
In a LFHD 
 approach (see, e.g. \cite{melo02} and references quoted therein),  one has
 \be
 n^{HD}_{VM}(k_{\perp})=
\int_0^{2\pi} d\theta_{ k_\perp}\int_0^1 {d\xi ~M^2_{free}\over
\xi (1-\xi)} ~ \times\nonu
|\psi_{VM}(\xi,{\bf k}_{\perp})|^2
 \ee
where $\psi_{VM}(\xi,{\bf k}_{\perp})$ is the eigenfunction of a relativistic
  mass
operator and $M^2_{free}=(m^2+|{\bf k}_{\perp}|^2)/\xi (1-\xi)$. In order to
implement our procedure for fixing  the free parameters in $\Lambda_{VM}$, 
we have first 
evaluated $n^{HD}_{VM}(k_{\perp})$
by adopting the eigenfunctions obtained by using the model mass operators of both
Refs. \cite{qcd_tob} and \cite{GI}, and then we have minimized the
difference $|n_{val}( k_{\perp})-n^{HD}_{VM}(k_{\perp})|$.
 It should be
pointed out that the VM
 spectra   are satisfactorily reproduced for each VM investigated
  in this contribution. As an example of  the reliability of the adopted
  LFHD models,  the masses of the considered neutral VMs 
  for the ground states  are compared with 
  the corresponding experimental
  values in Tables \ref{tab1} and  \ref{tab2} for the mass operators 
  of Refs. \cite{qcd_tob} and \cite{GI}, respectively. 
  Indeed,  also the masses of the
  first three/four (depending upon the considered VM) excited states are
  well reproduced by the model mass operators.

Once  the parameters have been determined, one can calculate both static (e.g., em decay rates)
and dynamical (e.g. parton distributions) quantities.  For completing  
this  section, devoted to the general presentation of the formalism, we give the
expression of 
the parton distribution,
i.e.
\be
q(\xi)= \int dk_\perp ~ k_\perp~f_1(\xi,k_\perp)\ee
in terms of the chiral-even transverse-momentum distribution,
$f_1(\xi,k_\perp)$, 
(see, e.g. \cite{fpps10} and \cite{pion09} for the pion case), that yields the 
 distribution of a constituent
inside the meson with the longitudinal and transverse components 
of the LF-momentum. It reads 
\be f_1(\xi,k_\perp)=~{N_c\over 4 P^+}~\int^{2\pi}_0 d\theta_{k_\perp}~
 \int\frac{dk^-}{(2\pi)^4}~\times\nonu
 ~{\bar\Lambda_{VM}(k,P )\over \left[(k-P )^2- m^2_2+\imath\epsilon\right]}
 {{ \bf T}^+_{\lambda,\lambda}(P,k)\over \left[(k-P)^2- m^2_2+\imath\epsilon\right]}
 \times \nonu{\Lambda_{VM}(k,P) \over (k^2 - m^2_1+\imath \epsilon) }
 \ee  
The explicit expression of  $f_1(\xi,k_\perp)$ will be presented elsewhere
\cite{fpps10}.
\section{THE MANDELSTAM FORMULA FOR THE EM DECAY CONSTANT}
In order to evaluate the em decay constant, $f_{VM}$, we adopt 
a Mandelstam-like formula \cite{mandel} (see also  \cite{pionprd} and
\cite{pion09}
for the pion case). 
The starting point is the {\em macroscopic} definition of $f_{VM}$, through 
the transition matrix element of the em current for a given neutral  VM, viz
\be
\langle 0|J^{\mu}(0)| P,\lambda \rangle =
\imath \sqrt{2}f_{VM} \epsilon_{\lambda}^{\mu}
\label{eq:fv} 
\ee
Let us remind that the decay constant $f_{VM}$ is related to the em decay width 
as follows
\be
\Gamma^{VM}_{e^+ e^-}=\frac{8\pi\alpha^2}{3}\frac{|f_{VM}|^2}{M_{VM}^3}
\label{eq:gamma} 
\ee
In our model, the transition matrix element in Eq. (\ref{eq:fv}) can be 
 approximated {\it microscopically}
{\it \`a la} Mandelstam through
\be
\langle 0|J^{\mu}(0)| P,\lambda \rangle=
{\cal Q}_{VM}~{{N_c}{\cal N} \over (2\pi)^4}\int d^4 k~\times \nonu
{\Lambda_{VM}(k,P,m_1,m_2)\over 
  (k^2-m^2_1+ \imath \epsilon)~[(P-k)^2-m^2_2
  + \imath \epsilon]} \times \nonu
{\rm Tr} [ {\cal V}_\lambda(P) ~(\psla k -\psla P +m_2)\gamma^\mu
(\psla k+m_1)]
\label{eq:mandel} 
\ee 
where $$ {\cal Q}_\rho= {(Q_u-Q_d) \over \sqrt{2}} \quad \quad
{\cal Q}_\phi= Q_s$$ $$
{\cal Q}_{J/\Psi}= Q_c \quad \quad
{\cal Q}_{\Upsilon}= Q_b
$$
with $Q_i$ the quark charge. The lengthy algebra for evaluating  the analytic 
integrations in Eq. (\ref{eq:mandel}) will be reported
elsewhere \cite{fpps10}.
\section{PRELIMINARY RESULTS}
In Figs. 1 and 2, the valence transverse-momentum distributions (cf Eq. (\ref{nk})) 
for 
$\rho$, $\phi$, $J/\psi$ and $\Upsilon$ are shown, for different sets of the free
parameters. 
In order to get some dynamical input, in Fig. 1
the free parameters of our Ansatz (see Eq. (\ref{bsa})) have been
determined through the model of Ref. \cite{qcd_tob},  while in Fig. 2
through the model of Ref. \cite{GI}. In the figures the distributions are
multiplied by a factor $k_\perp$,  to have an immediate intuition of the
transverse-momentum region relevant in the evaluation of the valence
probabilities. It is worth noting that the long tail of the heavy VM is an expected feature.
In Figs. 3 and 4, the parton distributions are presented. As it is shown, the  general 
behavior at the
end-points and the accumulation of the distribution for $x=1/2$ as
the CQ mass increases are recovered
by using our Ansatz.

Finally, in Tables 3 and 4, the preliminary values for the em decay widths, 
for the two
different choices of the free parameters,  
are compared
with the corresponding experimental data. It should be pointed out that while
the model of Ref. \cite{qcd_tob} yields reasonable results for the light sector
(let us remind that the confining interaction for the squared mass operator of
 \cite{qcd_tob} is parabolic),  in the heavy sector the model of Ref. \cite{GI}
 seems to behave better (for this model a linear confining interaction is considered
 in the mass operator). Such a comparison  suggests that an improved description
 of the confining interaction  for the squared mass operator  
  of Ref. \cite{qcd_tob} could lead the theoretical predictions of the em
  decays closer to the experimental ones.
\begin{table*}[tbh]
\caption{Preliminary VM  em decay widths, calculated using our Ansatz, 
(\ref{bsa}), and the
free parameters determined  through the comparison with the eigenfunctions of the
model \cite{qcd_tob} for the mass operator.}
\label{tab3}
\vspace{.3cm}
\hspace{3.5cm}\begin{tabular}{ccc} \hline
{VM} &   {$\Gamma_{e^+ e^-}^{th}$  
(keV)}& {$\Gamma_{e^+ e^-}^{exp}$  (keV)} \cite{PDG08}\\
\hline

$\rho$  &  8.579  & 7.04 $\pm$ 0.06  
\\ 
$\phi$ & 1.952  & 1.27 $\pm$ 0.04
\\  
$J/\psi$  & 2.526 & 5.55 $\pm$ 0.14   
\\  
$\Upsilon$ &  0.187 &1.340$\pm$ 0.018   
\\  \hline
\end{tabular}
 
\end{table*}

\begin{table*}[tbh]
\caption{The same as in table 1, but using   the eigenfunctions of the
model \cite{GI} to fix the free parameters.}
\label{tab4}
\vspace{.3cm}
\hspace{3.5cm}\begin{tabular}{ccc} \hline
{VM}   & {$\Gamma_{e^+ e^-}^{th}$  
(keV)}& {$\Gamma_{e^+ e^-}^{exp}$  (keV)} \cite{PDG08}\\
\hline

$\rho$  &  24.791  & 7.04 $\pm$ 0.06  
\\ 
$\phi$ & 4.342  & 1.27 $\pm$ 0.04
\\  
$J/\psi$  & 7.102 & 5.55 $\pm$ 0.14   
\\  
$\Upsilon$ & 0.493  &1.340$\pm$ 0.018   
\\  \hline
\end{tabular}
 
\end{table*}
\begin{figure}
  \vspace{.3cm} 
        {\includegraphics[width=7.5cm]{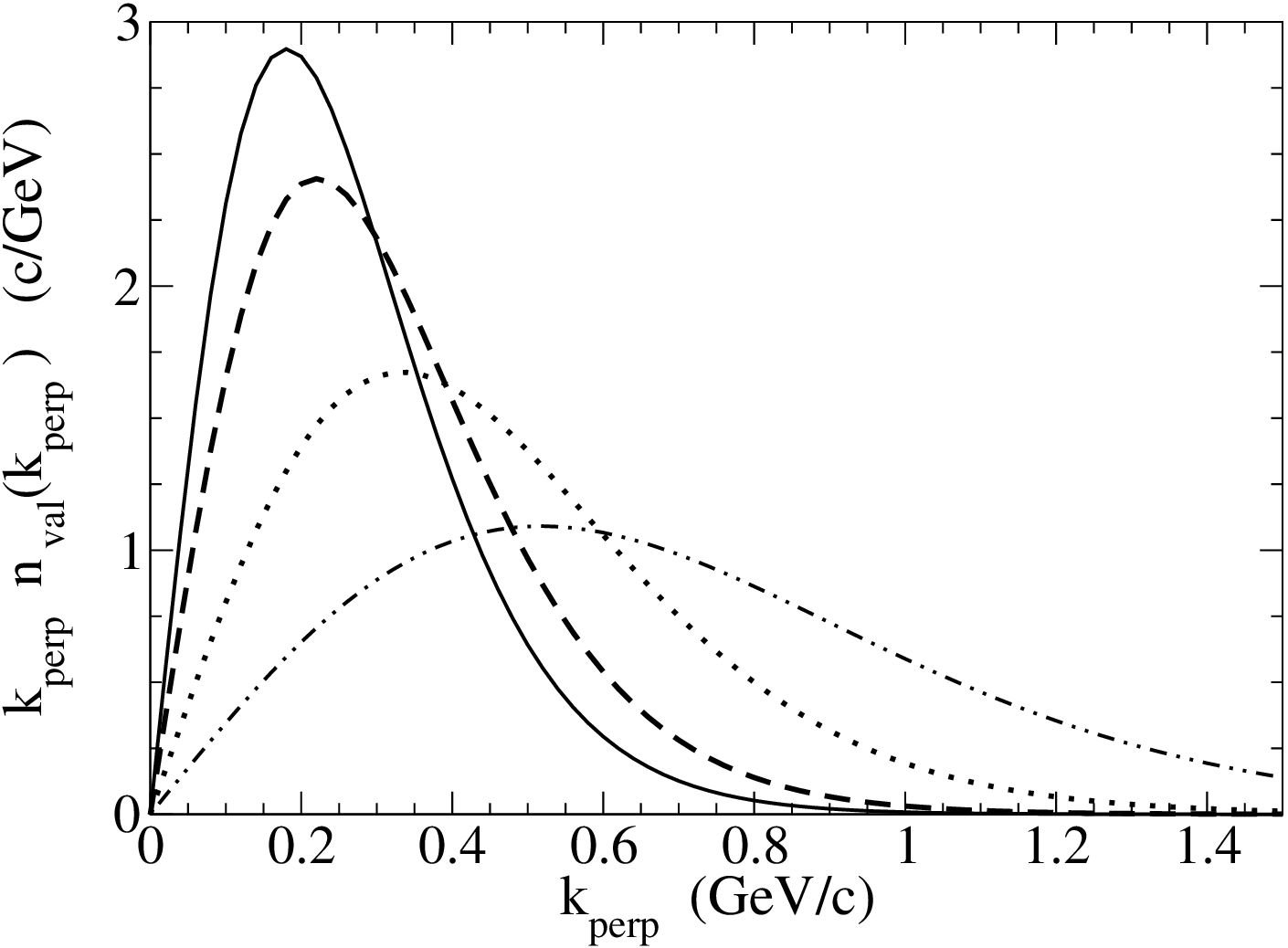}}

\caption{Valence transverse-momentum distributions for a constituent inside  
$\rho$,  $\phi$, $J/\Psi$ and $\Upsilon$
  vs the quark transverse momentum. The normalization is $
\int k_\perp~d{ k}_\perp ~n_{val}( k_{\perp})=1
$. Solid line: $\rho$; dashed line: $\phi$; dotted line: $J/\Psi$; 
double-dot-dashed line: $\Upsilon$. The
free parameters in (\ref{bsa}), have been determined  
through the comparison with the eigenfunctions of the
model \cite{qcd_tob}.} 
    
\end{figure}
\begin{figure}
   \vspace{-1.8cm}
        {\includegraphics[width=7.5cm]{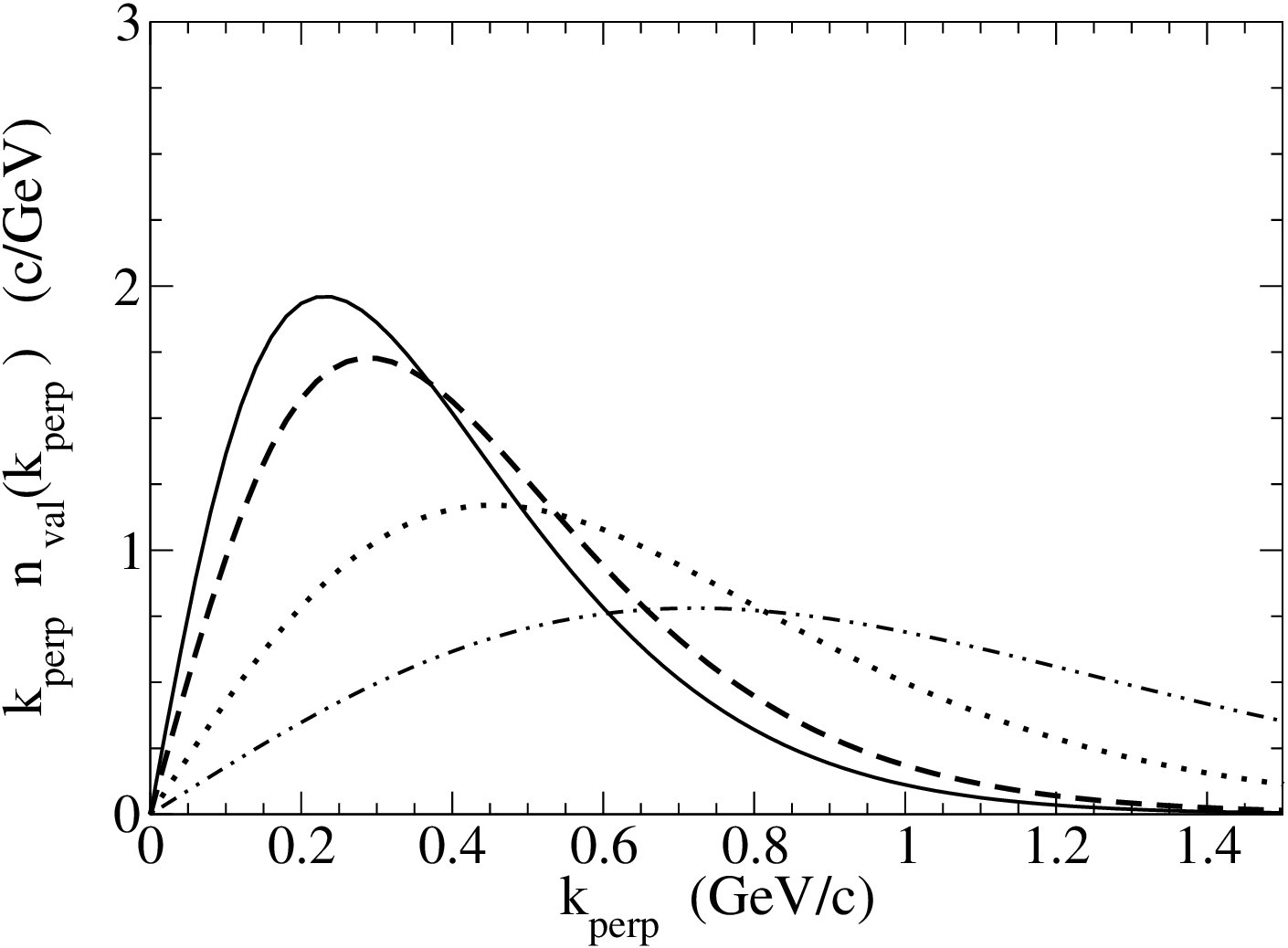}}

\caption{The same as in Fig. 1, but using   the eigenfunctions of the
model \cite{GI}  to fix the free parameters in (\ref{bsa}).}     
\end{figure}
\begin{figure}
 {\includegraphics[width=7.5cm]{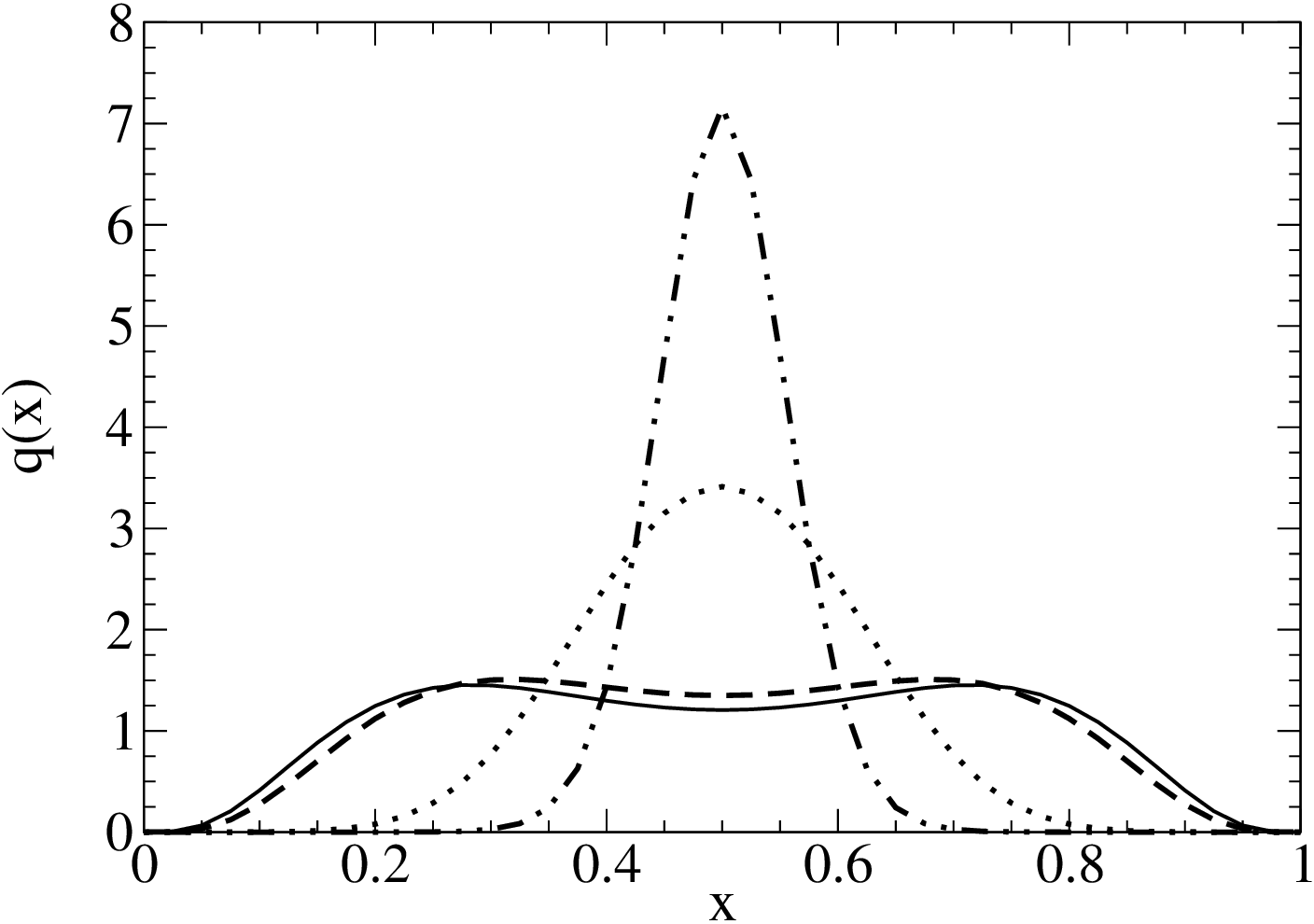}}   
\caption{Valence distributions for a constituent inside  
$\rho$,  $\phi$, $J/\Psi$ and $\Upsilon$
  vs $x$, with normalization $\int d\xi~q(\xi)=1$. The same line convention as in
  Fig. 1 is adopted. The
free parameters in (\ref{bsa}), have been determined  through 
the comparison with the eigenfunctions of the
model \cite{qcd_tob}.}
  \end{figure}
  \begin{figure}
 {\includegraphics[width=7.5cm]{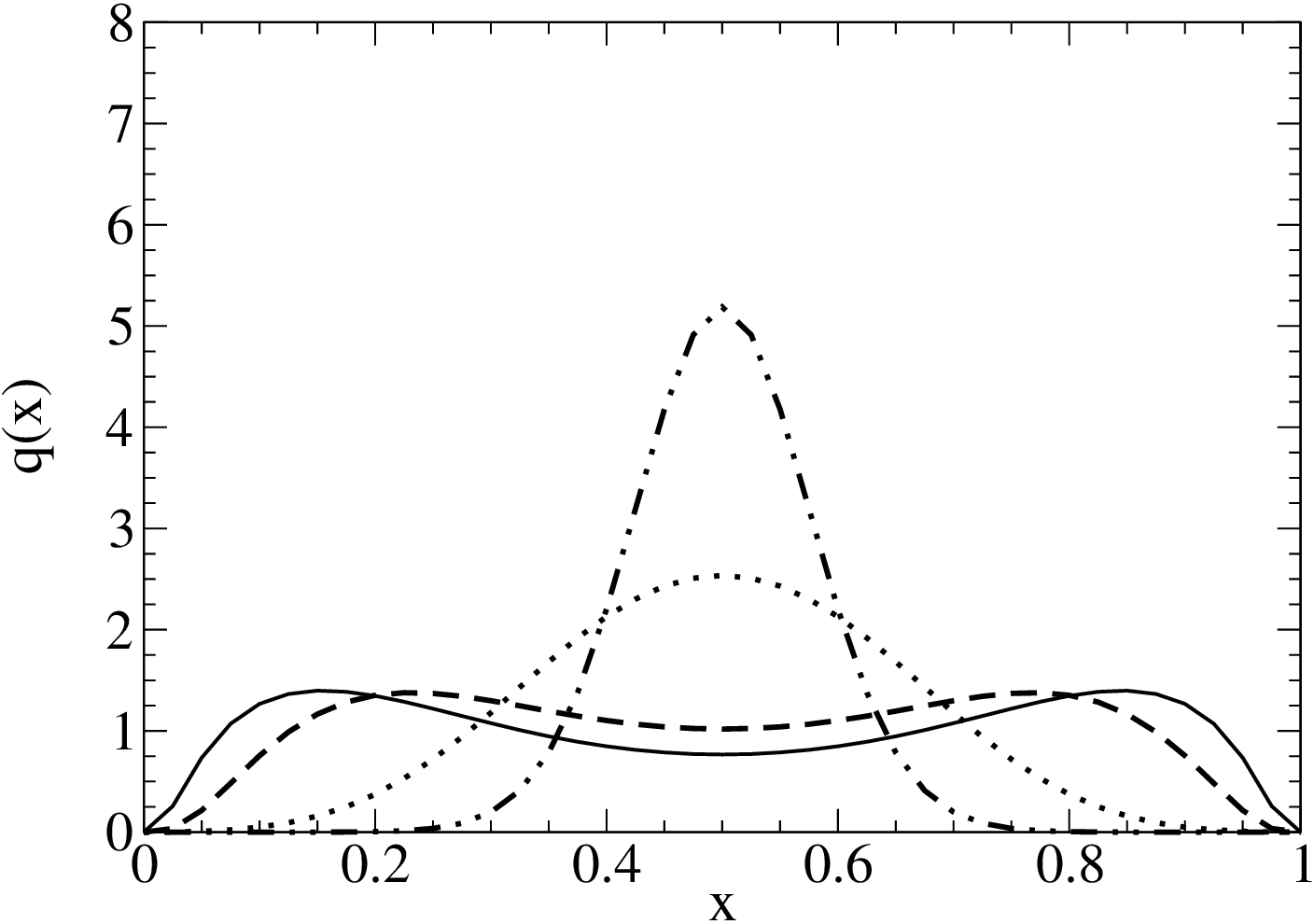}}   
\caption{The same as in Fig. 3, but using   the eigenfunctions of the
model \cite{GI} to fix the free parameters in (\ref{bsa}).}
  \end{figure}

\section{CONCLUSIONS}
 The preliminary results for
 some static and dynamical quantities for neutral VM, obtained
within our covariant description, both in the light and 
the heavy sectors, have been shortly presented. After
calculating
 the valence wave functions, Eq. (\ref{val}),  
and  the  transverse-momentum
distributions, Eq. (\ref{nk}), we have  fixed the three parameters in our Ansatz for the
Bethe-Salpeter amplitude through a comparison with the
transverse-momentum
distribution 
obtained within a Light-Front Hamiltonian Dynamics approach. The
eigenvectors of two  different mass operators, corresponding to  Ref.
\cite{qcd_tob} and Ref. \cite{GI}, have been considered in order to perform the comparison.
Then, we have evaluated the parton distributions and the em 
decay constants (cf
Table \ref{tab3} and Table \ref{tab4}).   

The work in progress will substantially improve the present calculations,
in two respect: both introducing a more refined Ansatz for the BS amplitude 
and considering new phenomenological mass operators.

This work was partially supported by the Brazilian agencies CNPq and
FAPESP and by Ministero della Ricerca Scientifica e Tecnologica.
 S.P. acknowledges the "Fondazione Della Riccia" for supporting
her research activity and   the hospitality of ITA. 

\end{document}